\RequirePackage{fix-cm}
\documentclass[smallextended]{svjour3}       

\usepackage{graphicx}
\usepackage{graphics}
\usepackage{amscd}
\usepackage{amsmath}
\usepackage{amsfonts}
\usepackage{amssymb}
\usepackage{enumerate}
\usepackage{float}
\restylefloat{table}

\usepackage{url}
\usepackage[hyperindex]{hyperref}
\newcommand{\F}{\mathbb{F}}
\newcommand{\Z}{\mathbb{Z}}

\smartqed  
\usepackage{graphicx}
\begin{document}

\title{The Discussion on Hulls of Cyclic Codes over the Ring $\Re =\Z_4+v\Z_4$, $v^2=v$
}


\author{Narendra Kumar         \and
        Abhay Kumar Singh
}

\institute{Department of Applied Mathematics \\
Indian Institute of Technology(ISM), Dhanbad, Jharkhand ,  826004, India\\
             \email{narendrakumar9670@gmail.com}         
         \\  and\\
    Department of Applied Mathematics \\
Indian Institute of Technology(ISM), Dhanbad, Jharkhand ,  826004, India\\
             \email{abhy@iitism.ac.in} 
}

\date{Received: date / Accepted: date}

\maketitle

\begin{abstract}
For odd length $n$, the  cyclic codes construction over $\Re= \Z_4[v]/ \langle v^2-v \rangle$ is provided. The hulls of cyclic codes over $\Re$ are studied. The average $2$-dimension $E(n)$ of the hulls of cyclic codes over $\Re$ is also conferred. Among these, the various examples of generators of hulls of cyclic codes over $\Re$ are provided, whose $\Z_4$-images are good $\Z_4$-linear codes with good parameters.
\end{abstract}

\section{Introduction}
\label{intro}
The study of hulls of cyclic codes have crucial importance because of their applications. Therefore the hulls of linear and cyclic codes over finite fields have been well studied. Assmus et al. \cite{15} introduced the theory of hull of linear codes to describe the finite projective planes. In this series, many authors discussed the properties of hulls to describe the complexity of some algorithms in \cite{18,19,20,21,22,23}. Again, Sendrier \cite{17} described the hulls of linear codes of length $n$ over square order field $\F_q$ and proved that  the average dimension of the hulls depend on the order of $\F_q$.  In \cite{24},Skersys discussed the average dimension $E_q(n)$ of the hulls of cyclic codes over finite fields and studied the character of $\dfrac{E_q(n)}{n}$, as $n$ tends to infinity. Later, Sangwisut et al. \cite{9} described  the generator polynomials and  dimensions of the hulls of cyclic and negacyclic codes over finite field. Recently, Jitman and Sangwisut \cite{7}  provided the average dimension
of the Hermitian hull of constacyclic codes over $\F_q$ and computed the upper and lower bounds. Furthermore, Jitman et al. \cite{16}  introduced the hulls of cyclic codes of odd length over $\Z_4$ and gave an algorithm to determine
types of hulls of cyclic codes over $\Z_4$. In this article, utilizing the results provided by Jitman et al. \cite{16}, we study the construction of hulls for our purpose.
\vskip 10pt
The cyclic codes are important family of linear codes.  The study of cyclic codes over finite rings have become a crucial topic of research. The cyclic codes over finite rings have been studied in a series of papers  \cite{4,5,11,26,27}. In particular, Hammons et al.\cite{27} done a brilliant work on codes over $\Z_4$ and discussed $\Z_4$ -linearity of different types of codes. After that many authors were attracted to do a lots of work over $\Z_4$. The ring $\Re= \Z_4[v]/ \langle v^2-v \rangle$ is an extension of ring $\Z_4$. For the first time Bandi and Bhaintwal \cite{13} introduced linear codes over finite ring $\Re$. They also studied the MacWilliams  identities for Lee weight and Gray enumerators for codes over $\Re$. After that, Gao et al. \cite{5} studied the different types of linear codes over $\Re$ and also obtained many new $\Z_4$-linear codes with good parameters. Recently, Dinh et al. \cite{11} introduced a new Gray map and they studied the all-constacyclic codes over $\Re$. Moreover, they also obtained the various new $\Z_4$-linear codes with better parameters. Further, Kumar and Singh \cite{12} studied the DNA computing by using cyclic codes over $\Re$. They provided various examples of reversible cyclic codes over $\Re$ and obtained many good $\Z_4$-linear codes with good parameters with the help of Gray map. These works encourage us to study the Hulls of cyclic codes over $\Re =\Z_4+vZ_4$, $v^2=v$.
\vskip 10pt
The remaining part of the article is organised in such manner. In section $2$, some basic properties and results related with cyclic codes over $\Re$ are conferred. Using the cyclic codes structure, the construction of hulls of cyclic codes over $\Re$ is studied in section $3$. The types of hulls of cyclic codes over $\Re$ are discussed in section $4$. Section $5$ is devoted to study of average $2$-dimension of cyclic codes over $\Re$.  In section $6$, various examples of hulls of cyclic codes over $\Re$ are given and we obtain the $\Z_4$-linear codes with good parameters according to $\Z_4$-database \cite{25}.
\section{Preliminaries}
As a linear code $C$ of length n over $\Z_4$ can be viewed as a vector space over $\F_2$ and
the concept of 2-dimension of C was introduced in \cite{28} and is given as $\dim_2(C)$ = log$_{2}(|C|)$.
Let $h = (h_0, h_1, \dots, h_{n-1}), k = (k_0, k_1, \dots, k_{n-1}) \in \Re^n$, the inner product is given as 
$$h \cdot k = h_0k_0 + h_1k_1 + \dots +h_{n-1}k_{n-1}.$$
 The dual code $C^{\perp}$ is conferred as
$$C^{\perp} = \{h \ | \ h \cdot k = 0, \forall k \in C \}.$$
Then $C$ is called {\it self-orthogonal} if  $C \subseteq C^{\perp}$ and $C$ is {\it self-dual} if $C = C^{\perp}$. The $hull$ of $C$ is defined as
\begin{equation*}
\text{Hull(C) }= C\cap C^{\bot}.
\end{equation*}

\vskip 5pt
Let $C$  be a cyclic code  over $\Z_4$ of odd length $n$ . Then generator of $C$ is given by

\begin{equation} C = \langle  f(x)g(x), 2 f(x)h(x)\rangle = \langle f(x)g(x), 2 f(x)\rangle ,
\end{equation} 
where $f(x),g(x),h(x)$ are unique monic polynomials over $\Z_{4}$ such that  $f(x)g(x)h(x)= x^n - 1$ (\cite{14}, Theorem $12.3.13$) and $|C|= 4^{\deg h(x)} 2^{\deg g(x)}$. Here, $C$ is said to be of type $4^{\deg h(x)} 2^{\deg g(x)}$. In this case, the $2$-dimension of $C$ is dim$_2(C)$= log$_2(|C|) = 2 \deg h(x) + \deg g(x)$.
A reciprocal polynomial of $h(x)= a_0+ a_1x+a_2 x^2+ \dots + x^r$ is defined to be $$ h^{*}(x)= a_0^{-1} x^{\deg h(x)} h\big(\dfrac{1}{x}\big).$$
Clearly, $(h^{*})^{*}(x)= h(x)$. A polynomial $h(x)$ is called self-reciprocal if $h^{*}(x)= h(x)$. From  (\cite{14}, Theorem 12.3.20), the generator of dual cyclic code $C^{\bot}$ is given as
\begin{equation}
\langle  h^{*}(x)g^{*}(x), 2 h^{*}(x)f^{*}(x)\rangle = \langle h^{*}(x)g^{*}(x), 2 h^{*}(x)\rangle
\end{equation} 

Some results are taken from \cite{16}, which are given below. 
Let $ord_{j}(i)$ denotes the mutiplicative order of $i$ modulo $j$, where $i$ and $j$ are coprime integers. Let $N_2: = \lbrace l \geq 1:l$ divides $ 2^i +1$ for some positive integer $i \rbrace$.  The factorization of $x^n-1$ is given as
\begin{equation}
x^n-1= \prod_{j|n, j\in N_2} \big( \prod_{i=1}^{\gamma(j)} g_{ij}(x) \big)  \prod_{j|n, j \notin  N_2} \big( \prod_{i=1}^{\beta(j)} f_{ij}(x)f_{ij}^{*}(x) \big) 
~~ \text{in} ~\Z_4[x] \end{equation}
where $$ \gamma(j) = \dfrac{\phi(j)}{ord_j(2)}, ~~ \beta(j) = \dfrac{\phi(j)}{2 ~ord_j(2)}$$ and $f_{ij}(x)$, $f_{ij}^{*}(x)$ form a monic basic irreducible reciprocal polynomial pair and $g_{ij}(x)$ is a monic basic irreducible self-reciprocal self-reciprocal polynomial. Let $ B_n= \deg \prod_{j|n, j\in N_2} \big( \prod_{i=1}^{\gamma(j)} g_{ij}(x) \big) $. Then \begin{equation} B_n = \deg \prod_{j|n, j\in N_2} \big( \prod_{i=1}^{\gamma(j)} g_{ij}(x) \big) = \sum_{j|n, j \in  N_2}\dfrac{\phi(j)}{ord_{j}(2)}ord_{j}(2) = \sum_{j|n, j \in  N_2} \phi(j).
\end{equation}
\vskip 5pt

The Gray map $\xi$ is taken from \cite{11} that transfers the elements of $\Re$ to elements of $\mathbb{Z}_{4}^2$ such that $$\xi(a+bv) = (a, 2b + a).$$  The Lee weight $w_{L}(a)$  in $\Z_{4}$ is given as $\min\lbrace{a, 4-a} \rbrace$. The Lee weight of any element of $\Re$ is defined as
$w_{L}(a+bv) = w_{L} (a) + w_{L} (2b + a)$. \vskip 5pt
\begin{theorem}\cite{11}
The Gray map $\xi$ is $\Z_{4}$-linear, and it is a distance-preserving map from $\Re^n$ (Lee distance) to $\Z^{2n}_{4}$ (Lee distance). 
\end{theorem}
\section{Hulls of Cyclic Codes over $\Re$}
In present section, the theory of hulls of cyclic codes over $\Re$ is discussed. First, we review some useful results, which will be used to describe  the hulls of cyclic codes over $\Re$.\vskip 10pt
 \noindent {\bf Theorem 3.1.}  \cite{16} Let $C$ be a cyclic code of odd length $n$ over $\Z_4$ generated by $\langle f(x) g(x), 2 f(x) \rangle$, where $x^n-1= f(x) g(x) h(x)$ and $f(x), g(x)$ and $ h(x)$ are pairwise coprime. Then Hull(C) is generated by 
$$
 \langle \text{lcm} (f(x)g(x), h^{*}(x)g^{*}(x)), 2 ~\text{lcm}( f(x), h^{*}(x)) \rangle
 $$ Furthermore, Hull(C) is of type $4^{\deg H(x)}2^{\deg G(x)}$, where $$H(x)= \text{gcd } (h(x), f^{*}(x)) ~~ \text{and} ~~ G(x)= \dfrac{x^n-1}{\text{gcd}(h(x), f^{*}(x))\cdot \text{lcm }(f(x), h^{*}(x))}. $$ 
 \vskip 10pt 
Some important results of cyclic codes over $\Re$ are given as follows.\vskip 10pt
 
\noindent {\bf Theorem 3.2.}  \cite{11}
 A linear code over $\Re$ is given as $C = vC_{1} \oplus (1+3v)C_{2}$. Then, $C$ is cyclic  if and only if $C_{1}$ and $C_{2}$ are cyclic over $\mathbb{Z}_{4}$. 	\vskip 10pt
 
\noindent {\bf Proposition 3.3.}  \cite{11}
Let n-tuple ${k}\in {\Re}^n$, then $ \xi \rho( {k}) = {\rho}^2 \xi ({k})$, where $\rho$ is the cyclic shift on $\mathbb Z_4^{2n}$. \vskip 10pt
 \noindent {\bf Proposition 3.4.}  \cite{11}
Let $C$ be a cyclic code of length $n$ over $\Re$. Then its Gray image $\xi(C)$ is a $2$-quasi-cyclic code of length $2n$ over $\mathbb{Z}_{4}$. \vskip 10pt
Let $C$ be cyclic code over $\Re$ of odd length $n$. By utilizing the  equation $1$, the generator of cyclic codes is described in next result.
 \vskip 10pt
 \noindent {\bf Theorem 3.5.}
Let  $C = v C_1 \oplus (1 - v) C_2$ be a cyclic code of odd length $n$ over $\Re$. Then 
\begin{center}
$C  = \langle v p_1 (x) q_1 (x) , 2 v p_1(x) , (1 - v)  p_2 (x) q_2 (x), 2 (1 - v) p_2 (x)\rangle, $
\end{center}
where $p_1 (x) q_1 (x) r_1 (x) = p_2 (x) q_2 (x) r_2 (x) = x^n-1$ and $C_1 = \langle p_1 (x) q_1 (x), 2 p_1 (x) \rangle$, $C_2 = \langle p_2 (x) q_2 (x), 2 p_2 (x)  \rangle$ over $\Z_{4}$, respectively. 

\begin{proof}
Let $\hat{C} = \langle v p_1 (x) q_1 (x) , 2 v p_1(x), (1 - v) p_2 (x) q_2 (x), 2 (1 - v) p_2 (x)  \rangle$, then it is obvious that $\hat{C} \subseteq C$. We have $ (v) C_1 = v \hat{C}$ since $v^2=v$ over $\Z_{4}$. Moreover, $(1 - v)^2 = (1 - v)$, then $(1 - v) C_2 = (1 - v) \hat{C}$. Thus, $ v C_1 \oplus (1-v) C_2 \subseteq \hat{C}$. Hence, $C=\hat{C}$ \end{proof}
 \vskip 10pt
\noindent {\bf Proposition 3.6.}\cite{11}
Let $C$ be a linear code of length $n$ over $\Re$, then $C^{\perp} = vC_1 ^{\perp} \oplus (1 - v)  C_2 ^{\perp}$.

\vskip 10pt
By using the dual cyclic codes $C^{\bot}$ over $\Z_4$ [given in equation 2], we describe the structure of dual of cyclic codes over $\Re$ in next result.

\vskip 10pt
\noindent {\bf Proposition 3.7.}
If  $C = \langle v p_1 (x) q_1 (x) , 2 v p_1(x) , (1 - v) p_2 (x) q_2 (x), 2 (1 - v) p_2 (x) \rangle$ is a cyclic code of odd length $n$ over $\Re$. Then $C^{\perp} =  \langle v r_1 ^* (x) q_1 ^* (x) , 2 v r_1 ^* (x), (1 - v) r_2 ^* (x) q_2 ^* (x), 2(1 - v) r_2 ^* (x) \rangle$, where $p^* (x) =  x^{\deg p(x)}$ $p(x^{-1})$. 
\vskip 10pt
\begin{proof}
If $C$ is a cyclic code, then dual $C^\perp$ is also cyclic code. Moreover, $C^{\perp} = v C_1 ^{\perp} \oplus (1-v)C_2 ^{\perp}$ by  Proposition $3.6$. Thus, equation $2$ and  Proposition $ 3.5$ imply our result. \end{proof}
\vskip 10pt 
\noindent {\bf Lemma 3.8.} Let $a=(a_0,a_1, \dots, a_{n-1})$ and $b=(b_0,b_1, \dots, b_{n-1})$ be vectors in $\Re^n$ with corresponding polynomials $a(x)$ and $b(x)$, respectively. Then $a$
is orthogonal to $b$ and all its shifts if and only if $a(x)b^{*}(x)=0 $ in $\Re[x]/ \langle x^n-1 \rangle$. 
\vskip 10pt
\begin{proof} Proof is similar to (\cite{14}, Theorem 12.3.18). \end{proof}
\vskip 10pt
Next result provide the generator of the hulls of the cyclic codes over $\Re$.
\vskip 10pt
\noindent {\bf Theorem 3.9.} Let $C$ be cyclic code of odd length $n$ over $\Re$ such that $C= vC_1 \oplus (1 - v) C_2$ and $C$ is generated by $$C = \langle v p_1 (x) q_1 (x) , 2 v p_1(x) , (1-v) p_2 (x) q_2 (x), 2 (1-v) p_2 (x)\rangle, $$ where $p_1 (x) q_1 (x) r_1 (x) = p_2 (x) q_2 (x) r_2 (x) = x^n-1$ and $C_1 = \langle p_1 (x) q_1 (x), 2 p_1 (x) \rangle$, $C_2 = \langle p_2 (x) q_2 (x), 2 p_2 (x)  \rangle$ over $\Z_{4}$, respectively.  Then Hull(C) has generator of the form 
$
 \langle v~ \text{lcm}~ (p_1 (x) q_1 (x), r^{*}_1 (x) q^{*}_1 (x)), 2v ~ \text{lcm}~( p_1 (x), r^{*}_1 (x)) , (1-v) ~ \text{lcm}~  (p_2 (x) q_2(x), r^{*}_2 (x) q^{*}_2 (x)), 2(1-v) ~ \text{lcm}~ ( p_2 (x), r^{*}_2(x))
\rangle 
$

\vskip 6pt In addition, Hull(C) is of the type $4^{\deg R_1(x)+\deg R_2(x)} 2^{\deg Q_1(x)+ \deg Q_2(x)}$, where $$ R_1(x) = \text{gcd} ~(r_1 (x), p^{*}_1 (x)), R_2(x) = \text{gcd} ~(r_2 (x), p^{*}_2 (x))$$ and $$Q_1(x) = \dfrac{x^n-1}{\text{gcd}~ ( r_1(x), p_1^{*}(x)). ~\text{lcm}~ (p_1(x), r_1^{*}(x))} ,  Q_2(x) = \dfrac{x^n-1}{\text{gcd} ~( r_2(x), p_2^{*}(x)). ~\text{lcm}~ (p_2(x), r_2^{*}(x))} .$$

\begin{proof} From Theorem $3.5$, the generator of a cyclic code $C$ over $\Re$ is conferred as
$$C = \langle v p_1 (x) q_1 (x) , 2 v p_1(x) , (1-v) p_2 (x) q_2 (x), 2 (1-v) p_2 (x)\rangle, $$ and from Proposition $3.7$, the dual of cyclic codes over $\Re$ is given as

$$C^{\perp} =  \langle v r_1 ^* (x) q_1 ^* (x) , 2 v r_1 ^* (x), (1 - v) r_2 ^* (x) q_2 ^* (x), 2(1 - v) r_2 ^* (x) \rangle$$
We consider $\bar{C}$ is a cyclic code over $\Re$, which has generator of the form 
$$\bar{C} = \langle v P_1 (x) Q_1 (x) , 2 v P_1(x), (1-v) P_2 (x) Q_2 (x), 2 (1-v) P_2 (x)\rangle, $$ where 
 $$ P_1(x) = \text{lcm} ~(p_1 (x), r^{*}_1 (x)), P_2(x) = \text{lcm} ~(p_2 (x), r^{*}_2 (x))$$  and $$Q_1(x) = \dfrac{x^n-1}{\text{gcd}~ ( r_1(x), p_1^{*}(x)). ~\text{lcm}~ (p_1(x), r_1^{*}(x))} ,  Q_2(x) = \dfrac{x^n-1}{\text{gcd} ~( r_2(x), p_2^{*}(x)). ~\text{lcm}~ (p_2(x), r_2^{*}(x))} .$$
$$ R_1(x) = \dfrac{x^n-1}{\text{lcm} ~( p_1(x)q_1(x),r^{*}_1(x)q^{*}_1(x) )}. = \text{gcd} ~(r_1 (x), p^{*}_1 (x)),$$
$$ R_2(x) = \dfrac{x^n-1}{\text{lcm} ~( p_2(x)q_2(x),r^{*}_2(x)q^{*}_2(x) )}. = \text{gcd} ~(r_2 (x), p^{*}_2 (x))$$
such that $x^n-1= P_1(x)Q_1(x)R_1(x)=P_2(x)Q_2(x)R_2(x)$ and $P_1(x),Q_1(x),R_1(x),P_2(x),Q_2(x),R_2(x)$ are pairwise coprime polynomials. Note that \vskip 5pt
$\langle v P_1 (x) Q_1 (x) , 2 v P_1(x) , (1-v) P_2 (x) Q_2 (x), 2 (1-v) P_2 (x)\rangle \subseteq  \langle v p_1 (x) q_1 (x) , 2 v p_1(x) , (1-v) p_2 (x) q_2 (x), 2 (1-v) p_2 (x)\rangle$ and also \vskip 5pt
$\langle v P_1 (x) Q_1 (x) , 2 v P_1(x), (1-v) P_2 (x) Q_2 (x), 2 (1-v) P_2 (x)\rangle \subseteq \langle v r_1 ^* (x) q_1 ^* (x) , 2 v r_1 ^* (x), (1 - v) r_2 ^* (x) q_2 ^* (x), 2(1 - v) r_2 ^* (x) \rangle,$ therefore we get $\bar{C} \subseteq \text{Hull(C)}$.
\vskip 5pt Now, we have to show that $\text{Hull(C)} \subseteq\bar{C}$.
Note that $\text{Hull(C)}$ is a cyclic code over $\Re$ and generated by
 $$\langle v L_1 (x) M_1 (x) , 2 v L_1(x), (1-v) L_2 (x) M_2 (x), 2 (1-v) L_2 (x)\rangle, $$
where the polynomials $ L_1 (x), M_1 (x), N_1 (x), M_2 (x), L_2 (x), N_2 (x)$ are pairwise coprime such that $$x^n -1 = L_1 (x) M_1 (x) N_1 (x)= M_2 (x) L_2 (x) N_2 (x).$$ As the $\text{Hull(C)} \subseteq C^{\perp}$ is orthogonal to $C$ and from Lemma $3.8$, we have 
$$ L_1 (x) M_1 (x) . 2 p^{*}_{1}(x)=0, ~~L_2 (x) M_2 (x) . 2 p^{*}_{2}(x)=0 $$ that means $ r^{*}_{1}(x)q^{*}_{1}(x)| L_1 (x) M_1 (x), ~r^{*}_{2}(x)q^{*}_{2}(x)| L_2 (x) M_2 (x)$ 
and also $$ 2L_1 (x). p^{*}_{1}(x)q^{*}_{1}(x)= 0, ~~2L_2 (x). p^{*}_{2}(x)q^{*}_{2}(x)= 0 $$ which provided that $r^{*}_{1}(x)|L_1 (x), r^{*}_{2}(x)|L_2 (x)$.
\vskip 5pt
The $Hull(C) \subseteq C$ is orthogonal to $C^{\perp}$ and from Lemma $3.8$, we have
$$ L_1 (x) M_1 (x). 2r_{1}(x)=0~ \text{and} ~L_2 (x) M_2 (x). 2r_{2}(x)=0, $$ that means $p_1 (x) q_1 (x)| L_1 (x) M_1 (x), ~ p_2 (x) q_2 (x)| L_2(x) M_2 (x)$ and also \\ $2 L_1 (x). r_1 (x) q_1 (x)= 0,~ 2 L_2 (x). r_2 (x) q_2(x)=0$, that gives $p_1 (x)|  L_1 (x) $ and $p_2 (x)|  L_2 (x) $. \vskip 5pt 
Therefore, $\text{lcm}( p_1(x)q_1(x), r^{*}_1(x)q^{*}_1(x))|L_1 (x) M_1 (x), \text{lcm}( p_2(x)q_2(x), r^{*}_2(x)q^{*}_2(x))|L_2 (x) M_2 (x)$ and  $\text{lcm}( r^{*}_1(x), p_1(x))|L_1(x),$ $~\text{lcm}( r^{*}_2(x), p_2(x))|L_2(x)$. \vskip 5pt
That means $ P_1(x)Q_1(x) | L_1 (x) M_1 (x), ~ P_2(x)Q_2(x)|L_2 (x) M_2 (x)$ and $ P_1(x) | L_1 (x), ~ P_2(x)|L_2 (x).$ Hence, $\text{Hull(C)} \subseteq \bar{C} $. Thus, $\text{Hull(C)} = \bar{C}$. \end{proof}
\section{\bf Types of Hulls of Cyclic Codes over $\Re$}
\vskip 10pt
Present section devotes to study the types of hulls of cyclic codes of odd length $n$ over $\Re$. For this, first we discuss some results.
\vskip 10pt
\noindent{\bf Lemma 4.1} \cite{16} Let $\beta$ be a positive integer. For $1 \leq i \leq \beta$, let $(v_i, z_i), (w_i, d_i) $ and $(u_i, b_i)$
be elements in $\lbrace (0, 0), (1, 0), (0, 1)\rbrace $. Let $a_i = \text{min}\lbrace 1-v_i- z_i,  w_i \rbrace + \text{min} \lbrace 1-w_i -d_i, v_i \rbrace.$
Then $a_i \in \lbrace 0, 1\rbrace$.  Moreover, the following statements hold.
\begin{enumerate}
\item $2-\text{min}\lbrace 1-v_i- z_i,  w_i \rbrace - \text{max} \lbrace v_i, 1-w_i -d_i \rbrace - \text{min} \lbrace 1-w_i -d_i, v_i \rbrace - \text{max} \lbrace w_i, 1-v_i - z_i \rbrace = z_i + d_i$. 
\item If $a_i = 0$, then $z_i + d_i \in \lbrace 0, 1, 2 \rbrace$.
\item  If $a_i = 1$, then $z_i + d_i =0 $.
\item Let $ a = \sum_{i=1}^{\beta} a_{ij},$ then $ \sum_{i=1}^{\beta} ( z_{i}+d_{i})= c$ for some $0 \leq c \leq 2(\beta -a)$. 
\end{enumerate}
\vskip 10pt 
\noindent{\bf Theorem 4.2}   The types of the hull of a cyclic code of
odd length n over $\Re$ are $4^{k_1}2^{k_2}$ , where
\begin{equation*}
k_1 = \sum\limits_{j|n, j \not \in N_2} ord_{j}(2). a_{1j} + \sum\limits_{j|n, j \not \in N_2} ord_{j}(2). a_{2j}~ \text{and}
 \end{equation*} 
 \begin{equation*}~ k_2 = \sum\limits_{j|n, j  \in N_2} ord_{j}(2). b_{1j} + \sum\limits_{j|n, j  \in N_2} ord_{j}(2). b_{2j} + \sum\limits_{j|n, j \not \in N_2} ord_{j}(2). c_{1j} + \sum\limits_{j|n, j \not \in N_2} ord_{j}(2). c_{2j},
\end{equation*}
$0 \leq a_{1j} ,a_{2j} \leq \beta (j), 0 \leq b_{1j} ,b_{2j} \leq \gamma (j)$ and $ 0 \leq c_{1j}  \leq 2( \beta (j)- a_{1j})$, $ 0 \leq c_{2j} \leq 2( \beta (j)- a_{2j})$.
\vskip 10pt
\begin{proof} From Theorem $3.9$,  $\text{Hull(C)}$ has type 
$$4^{\deg R_1(x)+\deg R_2(x)} 2^{\deg Q_1(x)+ \deg Q_2(x)}.$$ From equation $(3)$, we get $$p_{k}(x)= \prod\limits_{j|n, j\in N_2} \big( \prod\limits_{i=1}^{\gamma(j)} q_{kij}(x)^{u_{kij}} \big)  \prod\limits_{j|n, j \notin  N_2} \big( \prod\limits_{i=1}^{\beta(j)} p_{kij}(x)^{v_{kij}}p_{kij}^{*}(x)^{w_{kij}} \big)$$
$$q_{k}(x)= \prod\limits_{j|n, j\in N_2} \big( \prod\limits_{i=1}^{\gamma(j)} q_{kij}(x)^{b_{kij}} \big)  \prod\limits_{j|n, j \notin  N_2} \big( \prod\limits_{i=1}^{\beta(j)} p_{kij}(x)^{z_{kij}}p_{kij}^{*}(x)^{d_{kij}} \big)$$
$$r_{k}(x)= \prod\limits_{j|n, j\in N_2} \big( \prod\limits_{i=1}^{\gamma(j)} q_{kij}(x)^{1-u_{kij}- b_{kij}} \big)  \prod\limits_{j|n, j \notin  N_2} \big( \prod\limits_{i=1}^{\beta(j)} p_{kij}(x)^{1-v_{kij}-z_{kij}}p_{kij}^{*}(x)^{1-w_{kij}-d_{kij}} \big),$$
and 
$$p^{*}_{k}(x)= \prod\limits_{j|n, j\in N_2} \big( \prod\limits_{i=1}^{\gamma(j)} q_{kij}(x)^{u_{kij}} \big)  \prod\limits_{j|n, j \notin  N_2} \big( \prod\limits_{i=1}^{\beta(j)} p_{kij}(x)^{w_{kij}}p_{kij}^{*}(x)^{v_{kij}} \big)$$
$$q^{*}_{k}(x)= \prod\limits_{j|n, j\in N_2} \big( \prod\limits_{i=1}^{\gamma(j)} q_{kij}(x)^{b_{kij}} \big)  \prod\limits_{j|n, j \notin  N_2} \big( \prod\limits_{i=1}^{\beta(j)} p_{kij}(x)^{d_{kij}}p_{kij}^{*}(x)^{z_{kij}} \big)$$
$$r^{*}_{k}(x)= \prod\limits_{j|n, j\in N_2} \big( \prod\limits_{i=1}^{\gamma(j)} q_{kij}(x)^{1-u_{kij}- b_{kij}} \big)  \prod\limits_{j|n, j \notin  N_2} \big( \prod\limits_{i=1}^{\beta(j)} p_{kij}(x)^{1-w_{kij}-d_{kij}}p_{kij}^{*}(x)^{1-v_{kij}-z_{kij}}\big),$$ where $k=1,2$ and $(u_{kij},b_{kij}) ,(v_{kij},z_{kij}) ,(w_{kij},d_{kij}) \in \lbrace (0,0), (1,0) (0,1) \rbrace$. According to above relations, we have
\begin{align*}
R_{k}(x)&  = \text{gcd}(r_{k}(x), p^{*}_{k}(x)) \\
&= \prod\limits_{j|n, j\in N_2}  \prod\limits_{i=1}^{\gamma(j)} q_{kij}(x)^{\text{min} \lbrace 1-u_{kij}- b_{kij}, u_{kij} \rbrace}    \prod\limits_{j|n, j \notin  N_2}  \prod\limits_{i=1}^{\beta(j)} p_{kij}(x)^{\text{min} \lbrace 1-v_{kij}- z_{kij}, w_{kij} \rbrace}\\ & \times p_{kij}^{*}(x)^{\text{min} \lbrace 1-w_{kij}- d_{kij}, v_{kij} \rbrace} \\ &=  \prod\limits_{j|n, j \notin  N_2}  \prod\limits_{i=1}^{\beta(j)} p_{kij}(x)^{\text{min} \lbrace 1-v_{kij}- z_{kij}, w_{kij} \rbrace}p_{kij}^{*}(x)^{\text{min} \lbrace 1-w_{kij}- d_{kij}, v_{kij} \rbrace}\\
\end{align*}
that means 
\begin{align*}
\deg R_{k}(x)&  = \deg~ \text{gcd}(r_{k}(x), p^{*}_{k}(x)) 
\\ &= \deg ~ \prod\limits_{j|n, j \notin  N_2}  \prod\limits_{i=1}^{\beta(j)} p_{kij}(x)^{\text{min} \lbrace 1-v_{kij}- z_{kij}, w_{kij} \rbrace}p_{kij}^{*}(x)^{\text{min} \lbrace 1-w_{kij}- d_{kij}, v_{kij} \rbrace}
\\ & = \sum\limits_{j|n, j \notin  N_2} ord_{j}(2)\sum\limits_{i=1}^{\beta(j)} \big ( \text{min} \lbrace 1-v_{kij}-z_{kij}, w_{kij} \rbrace + \text{min} \lbrace 1- w_{kij}- d_{kij}, v_{kij} \rbrace \big ) \\
& = \sum\limits_{j|n, j \notin  N_2} ord_{j}(2)\sum\limits_{i=1}^{\beta(j)} a_{kij}, \text{where}~ 0 \leq a_{kij} \leq 1. \\ 
&= \sum\limits_{j|n, j \notin  N_2} ord_{j}(2) \cdot a_{kj},  \text{where}~ 0 \leq a_{kj} \leq \beta(j). \\
\end{align*}
In the similar way, we determine the $\deg Q_{k}(x)$, where $k= 1,2$. 
\begin{align*}
Q_{K}(x) & = \dfrac{x^n-1}{\text{gcd}~ ( R_k(x), p_k^{*}(x)). ~\text{lcm}~ (p_k(x), R_k^{*}(x))}\\ &=
\prod\limits_{j|n, j\in N_2}  \prod\limits_{i=1}^{\gamma(j)} q_{kij}(x)^{1-\text{max} \lbrace u_{kij}, 1-u_{kij}- b_{kij} \rbrace} \\&  \times  \prod\limits_{j|n, j \notin  N_2}  \prod\limits_{i=1}^{\beta(j)} p_{kij}(x)^{\text{min} \lbrace 1-v_{kij}- z_{kij}, w_{kij} - \text{max} \lbrace v_{kij}, 1-w_{kij}- d_{kij} \rbrace}\\&  \times p_{kij}^{*}(x)^{1-\text{min} \lbrace 1-w_{kij}- d_{kij}, v_{kij} \rbrace - \text{max} \lbrace w_{kij}, 1-v_{kij}- z_{kij} \rbrace } \\
\end{align*}
Thus,
\begin{align*}
\deg Q_{k}(x) & = \deg\prod\limits_{j|n, j\in N_2}  \prod\limits_{i=1}^{\gamma(j)} q_{kij}(x)^{1-\text{max} \lbrace u_{kij}, 1-u_{kij}- b_{kij} \rbrace} \\ &
  \times \prod\limits_{j|n, j \notin  N_2}  \prod\limits_{i=1}^{\beta(j)} p_{kij}(x)^{\text{min} \lbrace 1-v_{kij}- z_{kij}, w_{kij} - \text{max} \lbrace v_{kij}, 1-w_{kij}- d_{kij} \rbrace} \\ & \times p_{kij}^{*}(x)^{1-\text{min} \lbrace 1-w_{kij}- d_{kij}, v_{kij} \rbrace - \text{max} \lbrace w_{kij}, 1-v_{kij}- z_{kij} \rbrace } \\ & 
  = \sum\limits_{j|n, j \in  N_2} ord_{j}(2)\sum\limits_{i=1}^{\gamma(j)} \big ( 1- \text{max} \lbrace u_{kij},1-u_{kij}-b_{kij} \rbrace \big ) \\&  +  \sum\limits_{j|n, j \notin  N_2} ord_{j}(2)\sum\limits_{i=1}^{\beta(j)} \big (2-  \text{min} \lbrace 1- v_{kij}- z_{kij}, w_{kij} \rbrace \\ & - \text{max} \lbrace w_{kij},1-v_{kij}-z_{kij} \rbrace - \text{min} \lbrace 1- w_{kij}- d_{kij}, v_{kij} \rbrace \\&  -  \text{max} \lbrace v_{kij},1-w_{kij}-d_{kij} \rbrace \big ) \\
  &= 
  \sum\limits_{j|n, j \in  N_2} ord_{j}(2)\sum\limits_{i=1}^{\gamma(j)} \big ( 1- \text{max} \lbrace u_{kij},1-u_{kij}-b_{kij} \rbrace \big )\\&  + \sum\limits_{j|n, j \notin  N_2} ord_{j}(2)\sum\limits_{i=1}^{\beta(j)} \big( z_{kij}+d_{kij} \big) \\ & 
  = \sum\limits_{j|n, j \in  N_2} ord_{j}(2) \cdot b_{kj} + \sum\limits_{j|n, j \notin  N_2} ord_{j}(2)\cdot c_{kj} \\
\end{align*}
From Theorem $3.9$,  $\text{Hull(C)}$ is of the type 
$4^{\deg R_1(x)+\deg R_2(x)} 2^{\deg Q_1(x)+ \deg Q_2(x)}$, where
\begin{align*}
k_{1} & = \deg R_1(x) + \deg R_2(x) \\ & =  \sum\limits_{j|n, j \not \in N_2} ord_{j}(2). a_{1j} + \sum\limits_{j|n, j \not \in N_2} ord_{j}(2). a_{2j} \\ &
\text{and }\\ 
k_{2} & = \deg Q_1(x) + \deg Q_2(x) \\ & =  \sum\limits_{j|n, j \in  N_2} ord_{j}(2) \cdot b_{1j} + \sum\limits_{j|n, j \notin  N_2} ord_{j}(2)\cdot c_{1j}+ \sum\limits_{j|n, j \in  N_2} ord_{j}(2) \cdot b_{2j} + \sum\limits_{j|n, j \notin  N_2} ord_{j}(2)\cdot c_{2j},\\
 \end{align*}
where $0 \leq a_{1j} ,a_{2j} \leq \beta (j), 0 \leq b_{1j} ,b_{2j} \leq \gamma (j)$ and $ 0 \leq c_{1j}  \leq 2( \beta (j)- a_{1j})$, $ 0 \leq c_{2j} \leq 2( \beta (j)- a_{2j})$ \end{proof}
\vskip 10pt 
\noindent{ \bf Corollary 4.3} For odd length $n$ such that $n \in N_{2}$. Then the types of the hull of a cyclic codes over $\Re$ are of the form $4^{0}2^{k_{2}}$, where 
\begin{equation*}
k_{2} = \sum\limits_{j|n, j \in  N_2} ord_{j}(2) \cdot b_{1j} +\sum\limits_{j|n, j \in  N_2} ord_{j}(2) \cdot b_{2j}, ~~  0 \leq b_{1j} ,b_{2j} \leq \gamma (j)
\end{equation*}    \vskip 10pt
\begin{proof}
 From the Theorem $4.2$, if $n \in N_{2}$ and $j|n, j \in  N_2$, we get the required result. \end{proof} 
\vskip 10pt 
\begin{center}
{\bf Algorithm:} The types of the hull of a cyclic code of odd length $n$ over $\Re.$ \end{center} 
\begin{enumerate}
\item For each $j|n$, consider the following cases.
\begin{enumerate}
\item If $j \in N_2$, then compute $ord_j(2)$ and $\gamma( j)$.
\item If  $j \notin N_2$, then compute $ ord_j(2)$ and $\beta( j)$.
\end{enumerate}
\item Compute $$k_1 =  \sum\limits_{j|n, j \not \in N_2} ord_{j}(2). a_{1j} + \sum\limits_{j|n, j \not \in N_2} ord_{j}(2). a_{2j},$$ where $0 \leq a_{1j} ,a_{2j} \leq \beta (j)$.
\item Next, we determine $k_{2}$ for fixed $a_{1j}, a_{2j}$ in $2$, 
\begin{center}
$k_2 = \sum\limits_{j|n, j \in  N_2} ord_{j}(2) \cdot b_{1j} + \sum\limits_{j|n, j \notin  N_2} ord_{j}(2)\cdot c_{1j}+ \sum\limits_{j|n, j \in  N_2} ord_{j}(2) \cdot b_{2j} + \sum\limits_{j|n, j \notin  N_2} ord_{j}(2)\cdot c_{2j}$,

\end{center}
where $0 \leq b_{1j} ,b_{2j} \leq \gamma (j)$ and $ 0 \leq c_{1j}  \leq 2( \beta (j)- a_{1j})$, $ 0 \leq c_{2j} \leq 2( \beta (j)- a_{2j}).$
\end{enumerate}
\vskip 10pt
\begin{center} {\bf Example $1$: } \end{center}
For $n=7$, we discuss the types of hulls of cyclic codes over $\Re$. 
If $1\in N_2$, then $ord_1(2) = 1 $, and $\gamma(1)= 1$ and 
 if $7 \notin N_2$, then $ord_7(2) = 3$ and $\beta(7) =1$.
\vskip 5pt
Which implies
$k_{1}  = 3a_{17}+3 a_{27}$, with $0 \leq (a_{17}, a_{27}) \leq 1$. Then we have following cases.\vskip 5pt
\begin{enumerate}
\item
 Let $(a_{17}, a_{27}) = (0, 0)$, i.e.  $k_{1}=0$ and 

$k_{2} =  b_{11}+b_{21}+  3c_{17}+3 c_{27}$, with $0 \leq (b_{11}, b_{21}) \leq 1$ and $0 \leq c_{17}, c_{27} \leq 2$. 
\vskip 5pt
Then $k_{2} = (0,1,2,3,4,5,6,7,8,9,10,11,12,13,14)$ 
\item Let $(a_{17}, a_{27}) = (1, 0)$,  i.e. $k_{1}=3$ and 

$k_{2} =  b_{11}+b_{21}+  3c_{17}+3 c_{27}$, with $0 \leq (b_{11}, b_{21}) \leq 1$ and $c_{17} = 0, 0 \leq c_{27} \leq 2$.
\vskip 5pt
Then $k_{2} = (0,1,2,3,4,5,6,7,8)$ 
\item Let $(a_{17}, a_{27}) = (0, 1)$,  i.e. $k_{1}=3$ and 

$k_{2} =  b_{11}+b_{21}+  3c_{17}+3 c_{27}$, with $0 \leq (b_{11}, b_{21}) \leq 1$ and $c_{27} = 0, 0 \leq c_{17} \leq 2$.
\vskip 5pt
Then $k_{2} = (0,1,2,3,4,5,6,7,8)$
\item Let $(a_{17}, a_{27}) = (1, 1)$,  i.e. $k_{1}=6$ and 

$k_{2} =  b_{11}+b_{21}+  3c_{17}+3 c_{27}$, with $0 \leq (b_{11}, b_{21}) \leq 1$ and $c_{17},  c_{27} =0$.
\vskip 5pt
Then $k_{2} = (0,1,2)$
\end{enumerate}
 
\vskip 5pt

\begin{center}
 {\bf Example 2: }
 \end{center}
For $n=15$, we discuss the types of hulls of cyclic codes over $\Re$. 
 Let $1,3,5 \in N_2$, we get $ord_1(2) = 1,ord_3(2) = 2, ord_5(2) = 4 $, and $\gamma(1)=\gamma(3)=\gamma(5)= 1$ and  $15 \notin N_2$, then $ord_{15}(2) = 4$ and $\beta(15) =1$.
\vskip 5pt
Which implies
$k_{1}  = 4a_{115}+4 a_{215}$, with $0 \leq (a_{115}, a_{215}) \leq 1$. Then, we have following cases. \vskip 5pt
\begin{enumerate}
\item
 Let $ (a_{115}, a_{215}) = (0, 0)$, i.e.  $k_{1}=0$ and 

$k_{2} =  b_{11}+b_{21}+  2 b_{13} + 2 b_{23}+ 4b_{15}+4b_{25}+4c_{115}+4 c_{215}$, with $0 \leq (b_{11}, b_{21}, b_{13}, b_{23}, b_{15},b_{25}) \leq 1$ and $0 \leq c_{115}, c_{215} \leq 2$. 
\vskip 5pt
Then $k_{2} = (0,1,2,\dots, 30)$ 
\item Let $ (a_{115}, a_{215})= (1, 0)$,  i.e. $k_{1}=4$ and 

$k_{2} =    b_{11}+b_{21}+  2 b_{13} + 2 b_{23}+ 4b_{15}+4b_{25}+4c_{115}+4 c_{215}$, with $0 \leq (b_{11}, b_{21}, b_{13}, b_{23}, b_{15},b_{25}) \leq 1$ and $ c_{115}=0,~0 \leq c_{215} \leq 2$. 
\vskip 5pt
Then $k_{2} = (0,1,2, \dots, 22)$ 
\item  Let $ (a_{115}, a_{215})= (0, 1)$,  i.e. $k_{1}=4$ and 

$k_{2} =    b_{11}+b_{21}+  2 b_{13} + 2 b_{23}+ 4b_{15}+4b_{25}+4c_{115}+4 c_{215}$, with $0 \leq (b_{11}, b_{21}, b_{13}, b_{23}, b_{15},b_{25}) \leq 1$ and $0 \leq c_{115} \leq 2,  c_{215}=0$. 
\vskip 5pt
Then $k_{2} = (0,1,2, \dots, 22)$ 
\item Let $ (a_{115}, a_{215})= (1, 1)$,  i.e. $k_{1}=8$ and 

$k_{2} =    b_{11}+b_{21}+  2 b_{13} + 2 b_{23}+ 4b_{15}+4b_{25}+4c_{115}+4 c_{215}$, with $0 \leq (b_{11}, b_{21}, b_{13}, b_{23}, b_{15},b_{25}) \leq 1$ and $c_{115} = c_{215}=0$.
\vskip 5pt
Then $k_{2} = (0,1,2,\dots , 14)$
\end{enumerate}
\section{The Average $2$-Dimension $E(n)$}
\vskip 10pt In this section, the average $2$-dimension of the cyclic codes of length $n$ over $\Re$ is discussed. The average $2$-dimension of the cyclic codes over $\Z_{4}$ is given as 
\begin{equation}
E(n) = \sum\limits_{C \in C(n,4)} \dfrac{\deg_{2}(\text{Hull(C)}}{|C(n,4)|}, 
\end{equation} where $C(n,4)$ denotes the set of all cyclic codes over $\Z_{4}$.
\vskip 10pt
\noindent{ \bf Lemma 5.1}  \cite{16}
Let $(v,z), (w, d), (u,b) \in \lbrace (0,0),(1,0),(0,1) \rbrace$. Then 
\begin{enumerate}
\item $E(1- \text{max} \lbrace u, 1-u-b \rbrace) = \dfrac{1}{3}$
\item $E(2+ \text{min} \lbrace 1-v-z, w \rbrace - \text{max} \lbrace v, 1-w-d \rbrace + \text{min} \lbrace 1-w-d,v \rbrace - \text{max} \lbrace w, 1-v-z \rbrace ) = \dfrac{10}{9}$
\end{enumerate} 
Next, we study the formula of average $2$- dimension of cyclic codes over $\Re$ by utilizing above result and the expectation $E(Y)$, where $Y$ is the random variable of the $2$-dimension $dim _2(\text{Hull(C)})$ and $C$ is chosen randomly from $C(n,4)$ with uniform probability.
 \vskip 10pt
 \noindent{\bf Theorem 5.2} The value of $E(n) $ of hulls of cyclic codes of odd length $n$ over $\Re$ is conferred as
$$E(n) =\dfrac{ 10 n}{9}- \dfrac{4B_{n}}{9},$$
where $ B_{n}$ is given in Equation $(4)$.
\vskip 10pt
\begin{proof} The structure of cyclic codes $C$ of odd length $n$ over $\Re$ is determined as $$C = \langle v p_1 (x) q_1 (x) , 2 v p_1(x) , (1-v) p_2 (x) q_2 (x), 2 (1-v) p_2 (x)\rangle, $$ where $p_1 (x) q_1 (x) r_1 (x) = p_2 (x) q_2 (x) r_2 (x) = x^n-1$, where for $i= 1,2$, $p_i (x), q_i (x)$ and $ r_i (x)$ are pairwise coprime polynomials over $\Z_{4}$. The $\text{Hull(C)}$ has  type $$4^{\deg R_{1}(x) +\deg R_{2}(x)} 2^{\deg Q_{1}(x) +\deg Q_{2}(x)} $$ and  $2$- dimension of $\text{Hull(C)}$ is $$2(\deg R_{1}(x) +\deg R_{2}(x))+ \deg Q_{1}(x) +\deg Q_{2}(x)
$$ 
\begin{align*}
dim_{2}(\text{Hull(C))} & = 2(\deg R_{1}(x) +\deg R_{2}(x))+ \deg Q_{1}(x) +\deg Q_{2}(x) \\&
= 2 \bigg( \sum\limits_{j|n, j \notin  N_2} ord_{j}(2)\sum\limits_{i=1}^{\beta(j)} \bigg( \text{min} \lbrace 1-v_{1ij}-z_{1ij}, w_{1ij} \rbrace  + \text{min} \lbrace 1- w_{1ij}- d_{1ij}, v_{1ij} \rbrace \bigg) \\ &+ \sum\limits_{j|n, j \notin  N_2} ord_{j}(2)\sum\limits_{i=1}^{\beta(j)} \bigg( \text{min} \lbrace 1-v_{2ij}-z_{2ij}, w_{2ij} \rbrace   + \text{min} \lbrace 1- w_{2ij}- d_{2ij}, v_{2ij} \rbrace \bigg) \bigg) \\& + \sum\limits_{j|n, j \in  N_2} ord_{j}(2)\sum\limits_{i=1}^{\gamma(j)} \bigg( 1- \text{max} \lbrace u_{1ij},1-u_{1ij}-b_{1ij} \rbrace \bigg) \\ & +  \sum\limits_{j|n, j \notin  N_2} ord_{j}(2)\sum\limits_{i=1}^{\beta(j)} \bigg(2-  \text{min} \lbrace 1- v_{1ij}- z_{1ij}, w_{1ij} \rbrace  - \text{max} \lbrace w_{1ij},1-v_{1ij}-z_{1ij} \rbrace \\ & - \text{min} \lbrace 1- w_{1ij}- d_{1ij}, v_{1ij} \rbrace -  \text{max} \lbrace v_{1ij},1-w_{1ij}-d_{1ij} \rbrace \bigg)  \\& + \sum\limits_{j|n, j \in  N_2} ord_{j}(2)\sum\limits_{i=1}^{\gamma(j)} \bigg( 1- \text{max} \lbrace u_{2ij},1-u_{2ij}-b_{2ij} \rbrace \bigg) \\ & +  \sum\limits_{j|n, j \notin  N_2} ord_{j}(2)\sum\limits_{i=1}^{\beta(j)} \bigg(2-  \text{min} \lbrace 1- v_{2ij}- z_{2ij}, w_{2ij} \rbrace  - \text{max} \lbrace w_{2ij},1-v_{2ij}-z_{2ij} \rbrace \\ & - \text{min} \lbrace 1- w_{2ij}- d_{2ij}, v_{2ij} \rbrace -  \text{max} \lbrace v_{2ij},1-w_{2ij}-d_{2ij} \rbrace \bigg) \\ &
=   \sum\limits_{j|n, j \in  N_2} ord_{j}(2)\sum\limits_{i=1}^{\gamma(j)} \bigg( 1- \text{max} \lbrace u_{1ij},1-u_{1ij}-b_{1ij} \rbrace \bigg)+  \\ & \sum\limits_{j|n, j \notin  N_2} ord_{j}(2)\sum\limits_{i=1}^{\beta(j)} \bigg(2+ \text{min} \lbrace 1- v_{1ij}- z_{1ij}, w_{1ij} \rbrace  - \text{max} \lbrace v_{1ij},1-w_{1ij}-d_{1ij} \rbrace \\ & + \text{min} \lbrace 1- w_{1ij}- d_{1ij}, v_{1ij} \rbrace - \text{max} \lbrace w_{1ij},1-v_{1ij}-z_{1ij} \rbrace \bigg) \\ &
+  \sum\limits_{j|n, j \in  N_2} ord_{j}(2)\sum\limits_{i=1}^{\gamma(j)} \bigg( 1- \text{max} \lbrace u_{2ij},1-u_{2ij}-b_{2ij} \rbrace\bigg) \\ & +  \sum\limits_{j|n, j \notin  N_2} ord_{j}(2)\sum\limits_{i=1}^{\beta(j)} \bigg(2+ \text{min} \lbrace 1- v_{2ij}- z_{2ij}, w_{2ij} \rbrace  - \text{max} \lbrace v_{2ij},1-w_{2ij}-d_{2ij} \rbrace  \\ & + \text{min} \lbrace 1- w_{2ij}- d_{2ij}, v_{2ij} \rbrace  
 - \text{max} \lbrace w_{2ij},1-v_{2ij}-z_{2ij} \rbrace \bigg)
\end{align*}
Next, utilizing the Lemma $5.1$, we have
\begin{align*}
E(n)& = E(Y)\\ &
= E(2k_{1}+k_{2})\\&
= E(2 (~\deg R_1(x) +  \deg R_2(x)) + \deg Q_1(x)+\deg Q_2(x))\\&
=E \Bigg( \sum\limits_{j|n, j \in  N_2} ord_{j}(2)\sum\limits_{i=1}^{\gamma(j)} \bigg( 1- \text{max} \lbrace u_{1ij},1-u_{1ij}-b_{1ij} \rbrace \bigg) \Bigg) \\ & + E \Bigg( \sum\limits_{j|n, j \notin  N_2} ord_{j}(2)\sum\limits_{i=1}^{\beta(j)} \bigg(2+ \text{min} \lbrace 1- v_{1ij}- z_{1ij}, w_{1ij} \rbrace \\ & - \text{max} \lbrace v_{1ij},1-w_{1ij}-d_{1ij} \rbrace + \text{min} \lbrace 1- w_{1ij}- d_{1ij}, v_{1ij} \rbrace - \text{max} \lbrace w_{1ij},1-v_{1ij}-z_{1ij} \rbrace \bigg) \Bigg)\\ & + E \Bigg(\sum\limits_{j|n, j \in  N_2} ord_{j}(2)\sum\limits_{i=1}^{\gamma(j)} \bigg( 1- \text{max} \lbrace u_{2ij},1-u_{2ij}-b_{2ij} \rbrace \bigg) \Bigg) \\ & + E \Bigg(\sum\limits_{j|n, j \notin  N_2} ord_{j}(2)\sum\limits_{i=1}^{\beta(j)} \bigg(2+ \text{min} \lbrace 1- v_{2ij}- z_{2ij}, w_{2ij} \rbrace \\ & - \text{max} \lbrace v_{2ij},1-w_{2ij}-d_{2ij} \rbrace + \text{min} \lbrace 1- w_{2ij}- d_{2ij}, v_{2ij} \rbrace - \text{max} \lbrace w_{2ij},1-v_{2ij}-z_{2ij} \rbrace \bigg)\Bigg)\\ &
=  \sum\limits_{j|n, j \in  N_2} ord_{j}(2)\cdot \gamma(j) \cdot E \bigg( 1- \text{max} \lbrace u_{1ij},1-u_{1ij}-b_{1ij} \rbrace  \bigg) \\ & + \sum\limits_{j|n, j \notin  N_2} ord_{j}(2) \cdot \beta(j) \cdot \bigg( 2+ \text{min} \lbrace 1- v_{1ij}- z_{1ij}, w_{1ij} \rbrace \\ & - \text{max} \lbrace v_{1ij},1-w_{1ij}-d_{1ij} \rbrace + \text{min} \lbrace 1- w_{1ij}- d_{1ij}, v_{1ij} \rbrace - \text{max} \lbrace w_{1ij},1-v_{1ij}-z_{1ij} \rbrace \bigg)\\ &
+  \sum\limits_{j|n, j \in  N_2} ord_{j}(2)\cdot \gamma(j) \cdot E \bigg( \big( 1- \text{max} \lbrace u_{2ij},1-u_{2ij}-b_{2ij} \rbrace \big) \bigg) \\ & + \sum\limits_{j|n, j \notin  N_2} ord_{j}(2) \cdot \beta(j) \cdot \bigg( 2+ \text{min} \lbrace 1- v_{2ij}- z_{2ij}, w_{2ij} \rbrace \\ & - \text{max} \lbrace v_{2ij},1-w_{2ij}-d_{2ij} \rbrace + \text{min} \lbrace 1- w_{2ij}- d_{2ij}, v_{2ij} \rbrace - \text{max} \lbrace w_{2ij},1-v_{2ij}-z_{2ij} \rbrace \bigg)\\
& = \sum\limits_{j|n, j \in  N_2}  \phi(j)\cdot \dfrac{1}{3} + \sum\limits_{j|n, j \notin  N_2} \dfrac{\phi(j)}{2}\cdot \dfrac{10}{9}+ \sum\limits_{j|n, j \in  N_2}  \phi(j)\cdot \dfrac{1}{3} + \sum\limits_{j|n, j \notin  N_2}  \dfrac{\phi(j)}{2}\cdot \dfrac{10}{9} ~~\text{ using Lemma 5.1 } \\ &
= \dfrac{B_n}{3} + \dfrac{5(n-B_{n})}{9} +  \dfrac{B_n}{3} + \dfrac{5(n-B_{n})}{9} ~ ~\text{from equation 4} \\ & 
= \dfrac{10 n}{9} - \dfrac{4B_{n}}{9}\\ 
\end{align*} \end{proof}
\vskip 10pt 
\noindent{\bf Corollary 5.3} In the Theorem 5.2, we have $E(n)< \dfrac{10n}{9}$.
\vskip 6pt
 In Table $1$, we are given the average $2$-dimension $E(n)$ of the hull of cyclic codes of odd length $n$  from $55$ upto $151$. The $(\dagger)$ denotes the condition the $n \in N_{2}$ and otherwise $n \not \in N_{2}$ in the Table $1$. The Table $1$ is given after the references.

\begin{center}
\begin{center} {\bf Table $1$.} \end{center}
\begin{table}[ht]
\begin{tabular}{|p{1cm}|p{1cm}|p{3cm}|p{1cm}|p{1cm}|p{3cm}|}\hline
$n$ & $B(n)$ & $E(n)=\dfrac{10n-4B_{n}}{9}$ & $n$ & $B(n)$ & $E(n)=\dfrac{10n-4B_{n}}{9}$  \\[7pt]\hline
$55$ & $15$ & $\dfrac{490}{9}$ & $105$ & $13$ & $\dfrac{998}{9}$ \\ [7pt]\hline
${57}^{\dagger}$ & $21$ & $54$ & ${107}^{\dagger}$ & $107$ & $\dfrac{642}{9}$ \\ [7pt]\hline
${59}^{\dagger}$ & $59$ & $\dfrac{354}{9}$ & ${109}^{\dagger}$ & $109$ & $\dfrac{654}{9}$ \\[7pt]\hline
${61}^{\dagger}$ & $61$ & $\dfrac{366}{9}$ & $111$ & $39$ & $106$ \\[7pt]\hline
$63$ & $7$ & $\dfrac{602}{9}$ & ${113}^{\dagger}$ & ${113}$ & $\dfrac{678}{9}$  \\[7pt]\hline
${65}^{\dagger}$ & $17$ & $\dfrac{582}{9}$ & $115$ & $5$ & $\dfrac{1130}{9}$ \\[7pt]\hline
${67}^{\dagger}$ & $67$ & $\dfrac{402}{9}$ &  $117$ & $19$ & $\dfrac{1094}{9}$ \\[7pt]\hline
$69$ & $3$ & $\dfrac{678}{9}$ & $119$ & $17$ & $\dfrac{1122}{9}$ \\[7pt]\hline
$71$ & $1$ & $\dfrac{706}{9}$  & ${121}^{\dagger}$ & $111$ & $\dfrac{766}{9}$ \\[7pt]\hline
$73$ & $1$ & $\dfrac{726}{9}$ & $123$ & $43$ & $\dfrac{1158}{9}$ \\ [7pt]\hline
$75$ & $23$ & $\dfrac{658}{9}$ & ${125}^{\dagger}$ & $101$ & $\dfrac{846}{9}$ \\[7pt]\hline
 $77$ & $ 11$ & $ \dfrac{726}{9}$  & $127$ & $1$ & $\dfrac{1266}{9}$  \\[7pt]\hline
 $79$ & $1$ & $ \dfrac{726}{9} $  & ${129}^{\dagger}$ & $45$ & $\dfrac{1110}{9}$ \\[7pt]\hline
 ${81}^{\dagger}$ & $ 55$ & $ \dfrac{590}{9}$  & ${131}^{\dagger}$ & $131$ & $\dfrac{786}{9}$ \\[7pt]\hline
${83}^{\dagger}$ & $ 83$ & $ \dfrac{498}{9}$  & $133$ & $19$ & $\dfrac{1250}{9}$ \\[7pt]\hline
 $85$ & $ 21$ & $ \dfrac{766}{9}$  & $135$ & $23$ & $\dfrac{1258}{9}$ \\[7pt]\hline
$87$ & $ 31$ & $ \dfrac{746}{9}$  & ${137}^{\dagger}$ & $137$ & $\dfrac{822}{9}$ \\[7pt]\hline
$89$ & $ 1$ & $ \dfrac{886}{9}$  & ${139}^{\dagger}$ & $139$ & $\dfrac{834}{9}$ \\[7pt]\hline
$91$ & $ 13$ & $ \dfrac{858}{9}$  & $141$ & $3$ & $\dfrac{1398}{9}$ \\[7pt]\hline
$93$ & $ 3$ & $ 102$  &  $143$ & $23$ & $\dfrac{1338}{9}$ \\[7pt]\hline 
$95$ & $ 23$ & $ \dfrac{858}{9}$  & ${145}^{\dagger}$ & $33$ & $\dfrac{1318}{9}$ \\ [7pt] \hline
${97}^{\dagger}$ & $ 97$ & $ \dfrac{582}{9}$  & $147$ & $45$ & $\dfrac{1290}{9}$  \\ [7pt] \hline
${99}^{\dagger}$ & $ 17$ & $ \dfrac{922}{9}$  & ${149}^{\dagger}$ & $149$ & $\dfrac{894}{9}$ \\ [7pt] \hline
${101}^{\dagger}$ & $ 101$ & $ \dfrac{606}{9}$  & $151$ & $1$ & $\dfrac{1506}{9}$  \\ [7pt] \hline
$103$ & $1$ & $114$ & $153$ & $23$ & $\dfrac{1438}{9}$ \\[7pt] \hline
\end{tabular}
\end{table}
\end{center}
\section{Computational work}  
 
   In this section, we have provided the various examples of hulls of cyclic codes  of odd length over $\Re$. Among these, the generator polynomials of hulls of cyclic codes over $\Re$ are obtained. We know that cyclic codes over $\Re$ are mapped via Gray map into $\Z_{4}^2$ and these Gray images are $\Z_4$-linear codes. Thus, the Lee weights of the $\Z_{4}$-images are obtained. Various good $\Z_4$-linear codes with good parameters are obtained according to  data-base\cite{25}. Here, the monic polynomials are considered in ascending order such that  $x^4+3x^3+2x^2+x+1$ as $11231$ in Tables $2,3.$ The generator of hulls of cyclic codes over $\Re$ are given in following manner such that $(3+2v)+(1+2v)x+vx^2+ (2+v)x^3+2vx^4 = (31020)+\text{v}(22112)$ in Tables $2,3.$ The $({*})$ denotes good $\Z_4$-parameter.

   \vskip 10pt 
{\bf Example $3$}.  Let $n=15$ and $$
  x^{15} -1 = (x - 1)(x^2 + x + 1)(x^4 + x^3 + x^2 + x + 1)(x^4 + 
    3x^3 + 2x^2 + 1)(x^4 + 2x^2 + 3x + 1)$$ over $\Z_{4}$. In Table 2, the generator of hulls of cyclic codes over $\Re$ are obtained. 
 \begin{center}
 {\bf Table 2.} \vskip 10pt
\begin{tabular}{|p{2.4cm}|p{2.4cm}|p{2.4cm}|p{3.4cm}|p{2.2cm}|}

\hline
		 $\Z_4-\text{polynomials} $  & $\Z_4-\text{polynomials}$ & $\Z_4-\text{ polynomials}$ & ${\tiny \text{Generator of hulls over} ~\Re} $ & $\Z_{4}$-Parameters \\
		\hline
		 $p_1 = q_2=1131201$  & $q_1= r_2= 31$ & $r_1=p_2 = 102232311$ & $(220022200200000) + v (313301033221000)$ & $ (30, 4^0 2^{10}, 16)^{*}$ \\
		\hline
		 $p_1 = q_2= 102232311$  & $q_1= r_2= 111$ & $r_1=p_2 = 312321$ & $(200020220000000) + v (322312301111000)$ & $ (30, 4^0 2^{10}, 16)^{*}$ \\
		\hline
		 $p_1 = q_2= 13201$  & $q_1= r_2= 33300111$ & $r_1=p_2 = 10231$ & $(222202022002000) + v (300132303313000)$ & $ (30, 4^0 2^9, 2)^{*}$\\
			\hline
		 $p_1 = q_2= 10231$  & $q_1= r_2= 3001$ & $r_1=p_2 = 102232311$ & $(200220202222000) + v (313121231023000)$ & $ (30, 4^0 2^9, 12)^{*}$\\
		\hline
		 $p_1 = r_2= 10231$  & $r_1= q_2= 11111$ & $q_1=p_2 =3233221121$ & $(122110301311000) + v (122112101333000)$ & $(30, 4^4 2^4, 16)^{*}$\\
\hline
	$p_1 = r_2= 111$  & $q_1= p_2= 11111$ & $r_1=q_2 =3233221121$ & $(200222020000000) + v (022222000022000)$ & $(30, 4^0 2^9, 16)^{*}$\\
	\hline
	 $p_1 = r_2= 13201$  & $q_1= p_2= 10231$ & $r_1=q_2 =33300111$ & $(131101031201000) + v (113123211021000)$ & $(30, 4^4 2^7, 12)^{*}$\\
	\hline
	 $p_1 = r_2= 312321$  & $q_1= p_2=1021311$ & $r_1=q_2 =11111$ & $(102130123131000) + v (102132323111000)$ & $(30, 4^4 2^0, 24)^{*}$\\
	\hline
 $p_1 = r_2=  31 $  & $q_1= q_2=111$ & $r_1=p_2 =1001001001001$ & $(220220220220220) + v(000000000000000)$ & $(30, 4^0 2^2, 40)^{*}$\\
	\hline
	 $p_1 = r_2=1021211 $  & $q_1= q_2=113232201$ & $r_1=p_2 =31$ & $(133323233021000) + v (131321231023000)$ & $(30, 4^4 2^1, 20)$\\
	\hline
 $p_1 = r_2= 312321  $  & $q_1= q_2=102232311$ & $r_1=p_2 =111$ & $(122332121111000) + v (122132103333000)$ & $(30, 4^4 2^2, 16)^{*}$\\ 
	\hline
	 $p_1 = r_2= 3001  $  & $q_1= q_2=11111$ & $r_1=p_2 =130131031$ & $(220002200022000) + v (000000000000000)$ & $(30, 4^0 2^4, 24)^{*}$\\
	\hline
	 $p_1 = r_2= 11111$  & $q_1= q_2=3001$ & $r_1=p_2 =130131031$ & $(200200200200200) + v (000000000000000)$ & $(30, 4^0 2^3, 20)^{*}$\\
	\hline
	 $p_1 = r_2= 10231 $  & $q_1= q_2=13201$ & $r_1=p_2 =32200111$ & $(200220202222000) + v (022022220220000)$ & $(30, 4^0 2^4, 32)^{*}$\\
		\hline
	 $p_1 = r_2= (10231)(13201) $  & $q_1= p_2=(31)$ & $r_1=q_2 =(111)(11111) $ & $(202200220200000) + v (020022002022222)$ & $(30, 4^0 2^6, 24)^{*}$\\
\hline
\end{tabular}

  \end{center}
   \vskip 10pt 
{\bf Example $4$}.  Let $n=21$ and $
  x^{21} -1 =(x - 1)(x^2 + x + 1)(x^6 + 2x^5 + 3x^4 + 3x^2 + x + 
   1)(x^6 + x^5 + 3x^4 + 3x^2 + 2x + 1)(x^3 + 2*x^2 + x + 
   3)(x^3 + 3x^2 + 2x + 3)$ over $\Z_{4}$. In Table 3, the generator of hulls of cyclic codes over $\Re$ are obtained. 

 \begin{center}
{\tiny
 {\bf Table 3.} \vskip 10pt
\begin{tabular}{|p{2.8cm}|p{2.8cm}|p{2.8cm}|p{3.1cm}|p{1.6cm}|}

\hline
		 $\text{ $\Z_{4}$-polynomials} $  & $\text{ $\Z_{4}$-polynomials}$ & $\text{$\Z_{4}$- polynomials}$ & $\text{Generator of hulls over}~ \Re$ & $\Z_{4}$-Parameters \\
		\hline
		 $p_1 = q_2=1322121$  & $q_1= r_2= 1301301031031$ & $r_1=p_2 = 3231$ & $(222202220022020200000) + v (121332032333021031100)$ & $ (42, 4^{0}2^{9}, 24)^{*}$ \\
		\hline
		 $p_1 = q_2= 132120121$  & $q_1= r_2= 30000001$ & $r_1=p_2 = 1130321$ & $(202022220202220000000) + v (100120102002320000000)$ & $ (42, 4^{0} 2^{28}, 4)$ \\
			\hline
		 $p_1 = r_2= 333222333111$  & $q_1= p_2= 3001002001$ & $r_1=q_2 = 31$ & $(300022322300122000000) + v (300222322100122000000)$ & $(42, 4^{9} 2^{1}, 12)$ \\
			\hline
		 $p_1 = q_2=3231$  & $q_1= r_2= 3100000310000031$ & $r_1=p_2 = 3121$ & $(300022322300122000000) + v (300222322100122000000)$ & $(42, 4^{0} 2^{3}, 24)^{*}$ \\
		\hline 
		 $p_1 = q_2= 1230311$  & $q_1= r_2=  3330000111$ & $r_1=p_2 = 1130321$ & $(000000000000000000000) + v (312320332212012100000)$ & $(42, 4^{0} 2^{6}, 16)^{*}$ \\
			\hline 
	 $p_1 = q_2= 3001$  & $q_1= p_2=  3001002001$ & $r_1=r_2 =3002003001$ & $(300000300300100000000) + v (300000300300100000000)$ & $(42, 4^{10} 2^{3}, 2)$ \\
		\hline 
		 $p_1 = q_2=12311$  & $q_1= r_2= 111$ & $r_1=p_2 =3311211102331321$ & $(222202220022020200000) + v (123132030331221011100)$ & $(42, 4^0 2^{27}, 2)^{*}$ \\
		
			\hline 
		 $p_1 = r_2=  111 $  & $q_1= p_2= 3121$ & $r_1=q_2 =10203100123013111$ & $(200022000000000000000) + v (002202020222002022200)$ & $(42, 4^{0} 2^{16}, 12)^{*}$ \\
		\hline 
		 $p_1 = q_2= 1233031201$  & $q_1= r_2=111 $ & $r_1=p_2 =10213030321$ & $(202000022002000000000) + v (132103000313100000000)$ & $(42, 4^{0} 2^{19}, 16)^{*}$ \\
			\hline 
		 $p_1 = r_2= 302031$  & $q_1= q_2=31$ & $r_1=p_2 =3311211102331321$ & $(311010031101003110100) + v (032300003230000323000)$ & $(42, 4^3 2^4, 24)^{*}$ \\
		\hline 
		 $p_1 = r_2=13031022031 $  & $q_1= q_2=111$ & $r_1=p_2 =3002003001$ & $(320100120020100000000) + v (000320220200000000000)$ & $(42, 4^9 2^{23}, 2)^{*}$ \\
	
		\hline
		 $p_1 = r_2=1212231$  & $q_1= q_2=1230311 $ & $r_1=p_2 = 3002003001$ & $(323310212311201231100) + v (030300003210000301000)$ & $(42, 4^3 2^9, 16)$ \\ 
			\hline
		 $p_1 = r_2=1001001001001001001$  & $q_1= q_2= 111$ & $r_1=p_2 = 31$ & $(323310212311201231100) + v (030300003210000301000)$ & $(42, 4^0 2^2, 56)$ \\
		\hline
			 $p_1 = q_2=123333321$  & $q_1= r_2= 1301301031031$ & $r_1=p_2 = 31$ & $(202202222022020000000) + v (020202200222020000000)$ & $(42, 4^0 2^8, 24)^{*}$ \\ 
			\hline
			 $p_1 = p_2=10213030321$  & $q_1= r_2= 121021231$ & $r_1=q_2 = 3231$ & $(130120312010212100000) + v (202032303021233001100)$ & $(42, 4^6 2^9, 12)$\\
			\hline
		
\end{tabular}
}
  \end{center}
 
  \section{Conclusion} \vskip 6pt
   In this article, the construction of cyclic codes of odd length over $\Re$ is conferred. The generators of hulls of  cyclic codes over $\Re$ are studied. Moreover, the types of hulls of cyclic codes are discussed. Further, the average $2$-Dimension $E(n) $ are also conferred. Among, these many new $\Z_4$-linear codes are obtained, which have good parameters. The discussion of hulls of cyclic codes of even length over $\Re$ will be another open problem.



\begin{thebibliography}{}
%
%
\bibitem{1} T. Abualrub,  R. Oehmke, On the Generators of $\Z_4$ Cyclic Codes of Length $2^e$, \emph{IEEE Transactions on Information Theory}, ${\bf 49}, 2126-2133({2003})$ .

\bibitem{2} M. Ozena, F. Z. Uzekmeka,  N. Aydin and N. T. Özzaima, Cyclic and some constacyclic codes over the ring $\Z_{4}[u]/\langle u^2-1 \rangle$, \emph{Finite Fields Appl.}, ${\bf 38}, 27-39(2016)$. 



\bibitem{3} V. S. Pless, P. Sol\'{e} and Z. Qian, Cyclic self-dual $\Z_{4}$-codes, \emph{Finite Fields and Their Applications},$ {\bf 3}(1997),48-69$.

\bibitem{4} B. Yildiz, and N.Aydin, On cyclic codes over $\Z_4+u\Z_4$ and their $\Z_4$-images, \emph{Int. J.  Information and Coding Theory}, ${\bf 2(4)}, 226-237(2014)$.

\bibitem{5} J. Gao, F. W. Fu  and Y. Gao, Some classes of linear codes over $\Z4 + v\Z4$ and their
applications to construct good and new $\Z_{4}$-linear codes, \emph{Applicable Algebra in Engineering, Communication and Computing},$ 1 - 23 (2016)$.
\bibitem{6} S. Jitman and E. Sangwisut, (2017). The Average Dimension of the Hermitian Hull of Constayclic Codes over Finite Fields. {\em arXiv preprint arXiv:1702.00275}.
\bibitem{7} S. Jitman and E. Sangwisut, The average dimension of the Hermitian hull of cyclic codes over finite fields of square order, in: AIP  Proceedings of ICoMEIA 2016, 1775  Article ID 030026 (2016).
\bibitem{8} J. Leon, Permutation group algorithms based on partition, I: theory and algorithms, \emph{J. Symbolic Comput.}, ${\bf 12}, 533-583(1991) $.
\bibitem{9} E. Sangwisut, S. Jitman, S. Ling and P. Udomkavanich, Hulls of cyclic and negacyclic
codes over finite fields, \emph{Finite Fields Appl.}, ${\bf 33}, 232-257, (2015) $.

   \bibitem{11}  H.Q. Dinh, A.K. Singh, N. Kumar and S. Sriboonchitta, On constacyclic codes over {$\Z_4[v]/\langle v^2-v \rangle$} and their Gray images,  \emph{IEEE Communications Letters}, ${\bf 22(9)}, 1758-1761(2018)$.
\bibitem{12} N. Kumar  and A.K. Singh, DNA computing over the ring {$\Z_4[v]/\langle v^2-v \rangle$}, \emph{ International Journal of Biomathematics},  ${\bf 11(7)} , 1850090(2018)$.
\bibitem{13} R. K. Bandi  and  M. Bhaintwal, Codes over $\mathbb{Z}_{4} + v\mathbb{Z}_{4}$,  Advances in Computing, Communications and Informatics (ICACCI, 2014 International Conference on. $422 - 427(2014)$.
\bibitem{14}  W. C. Huffman, and  V. Pless,  \emph{Fundamentals of error-correcting codes}. Cambridge university press, (2010).
\bibitem{15}
E. F. Assmus and J. D. Key, Affine and projective planes, \emph{Discrete Math.}, ${\bf 83},
161-187(1990)$.
\bibitem{16}
S.~Jitman, E.~Sangwisut, and P.~Udomkavanich, ``Hulls of cyclic codes over
  $\mathbb{Z}_4$,'' {\em arXiv preprint arXiv:1806.07590}, (2018).
\bibitem{17} N. Sendrier On the dimension of the hull. \emph{SIAM Journal on Discrete Mathematics}, ${\bf 10(2)}, 282-293 (1997)$.
\bibitem{18} J.S. Leon,  An algorithm for computing the automorphism group of a Hadamard matrix, \emph{J. Comb. Theory}, ${\bf 27}(3),  289-306 (1979)$.
\bibitem{19} J.S. Leon, Computing automorphism groups of error-correcting codes, \emph{IEEE Trans. Inf. Theory}, ${\bf 28(3)}, 496-511(1982).$
\bibitem{20} J.S. Leon, Permutation group algorithms based on partition, I: theory and algorithms, \emph{J. Symb. Comput.}, ${\bf 12(4-5)}, 533-583 (1982)$.
\bibitem{21} J.S. Leon, Partitions, refinements, and permutation group computation, in: DIMACS Ser., \emph{Discrete Math. Theoret. Comput. Sci.}, $ {\bf 28} , 123-158 (1997)$.
\bibitem{22} N. Sendrier , Finding the permutation between equivalent codes: the support splitting algorithm, \emph{IEEE Trans. Inf. Theory}, ${\bf 46(4)} ,1193-1203(2000)$.
\bibitem{23} N. Sendrier and G. Skersys, On the computation of the automorphism group of a linear code, in: Proceedings of \emph{IEEE ISIT’2001, Washington}, DC, $13{2001}$.
\bibitem{24} G. Skersys,
The average dimension of the hull of cyclic
codes
\emph{ Discrete applied mathematics}, $ {\bf 128(1)}  , 275-292(2003)$.
\bibitem{25} Database of $\Z_{4}$ codes (on line), {\tt http://www.$\Z_{4}$codes.info} (accessed on $3$ September $(2017)$).
\bibitem{213} Aydin N., Asamov T., A database of $\Z_{4}$ codes, {\it J. Comb. Inf. Syst. Sci.} ${\bf 34}, 1-12,(2009)$.
\bibitem{26}  A.K. Singh and P.K. Kewat, On cyclic codes over the ring $\mathbb{Z} _{p} [u]/ {\langle u^ k \rangle}$, {\it Designs, Codes and Cryptography} ${\bf 74}  1-13(2015),$.
\bibitem{27} R. A. Hammons, V. P. Kumar, A. R. Calderbank, N. Sloane and Sole, P., The $\Z_4$-linearity of Kerdock, Preparata, Goethals, and related codes. \emph{IEEE Transactions on Information Theory}, $ {\bf 40}(2), 301-319(1994)$.
\bibitem{28}
 V. V. Vazirani, H. Saran and B. S. Rajan, An efficient algorithm for constructing
minimal trellises for codes over finite abelian groups, \emph{IEEE Trans. Inform.
Theory}, $ {\bf 42}, 839-1854(1996)$.

\end{thebibliography}


\end{document}